\documentstyle[12pt]{article}
\input epsf

\begin{document}

\vspace*{3cm}
\begin{center}
{\Large {\bf STRUCTURE FORMATION IN INFLATIONARY COSMOLOGIES}}\\
\vspace*{1cm}
{\large Andrew R.~Liddle} \\
{\em Astronomy Centre, University of Sussex, Falmer, Brighton BN1 
9QH,~~~U.~K.}

\vspace*{20pt}
{\bf Abstract}
\end{center}

\noindent
A brief account is given of large-scale structure modelling based on the 
assumption that the initial perturbations arise from inflation. A recap is 
made of the implications of inflation for large-scale structure; under the 
widely applicable slow-roll paradigm inflation adds precisely two extra 
parameters to the normal scenarios, which can be taken to be the tilt of the 
density perturbation spectrum and the amplitude of gravitational waves. Some 
comments are made about the {\it COBE} normalization. A short description is 
given of an analysis combining cosmic microwave background anisotropy data 
and large-scale structure data to constrain cosmological parameters, and the 
case of cold dark matter models with a cosmological constant is used as a 
specific illustration.

\vspace*{40pt}
\begin{center}
{\em To appear, Proceedings of the Moriond Conference on Microwave 
Background Anisotropies, March 16th--23rd 1996}\\
\vspace*{20pt}
Sussex preprint SUSSEX-AST 96/4-2; astro-ph/9604105
\end{center}

%%%%%%%%%%%%%%%%%%%%%%%%%%%%%%%%%%%%%%%%%%%%%%%%%%%%%%%%%%%%%%%%%%%%%%

\newpage
\section{Introduction}

In the medium term, the cosmic microwave background (CMB) shall undoubtedly 
prove an extremely powerful tool in constraining cosmological parameters. 
Already, the {\it COBE} observations of large-angle anisotropies provide the 
most accurate and unambiguous constraint on the spectra of perturbations in 
the universe. However, at the present time the best route to constraining 
cosmological parameters is not through the CMB alone, but from the 
combination of microwave data with a large number of measures of the power 
spectrum from large-scale structure observations.

In recent work, my collaborators and I have sought to test a wide parameter 
space of large-scale structure models, using linear and quasi-linear theory. 
We have written three papers on this topic, covering cold dark matter 
(CDM) models in open universes$^{1)}$ and in flat universes with a 
cosmological constant$^{3)}$, and the case of a critical density 
universe$^{2)}$ where we also allow a fraction of the dark matter to be 
hot. The key ingredient of our work is to take the inflationary hypothesis 
seriously, and to take advantage of the extra parameters that slow-roll 
inflation lends to large-scale structure modelling.

There isn't space here to give the full details of this work, so instead 
I'll concentrate on a couple of aspects and illustrate the outcome by 
showing some results from our investigation of CDM models in flat 
universes$^{3)}$.

%\section{Parameters for large-scale structure}

\section{Inflationary parameters}

The simplest approximation for the production of perturbations from 
inflation, that it gives a scale-invariant density perturbation spectrum 
and nothing else, is woefully inadequate to describe the output of most 
inflation models. It is necessary to make a better approximation, which 
acknowledges that as well as producing density perturbations inflation will 
produce a spectrum of gravitational waves. Based on the slow-roll 
approximation, one finds that inflation predicts power-law spectra of 
both density perturbations and gravitational waves. This gives a total of 
four parameters, two amplitudes and two spectral indices. This turns out to 
be an excellent approximation for almost all known inflationary models, and 
may well be all that is ever needed. 

Interestingly, these four parameters are not all independent. The reason is 
that we have extracted two continuous functions, the spectra, from a single 
input function, the inflaton potential. The only way you can get two 
continuous functions from one is if they are related, and at this level of 
approximation it turns out that one of the four parameters is redundant. 
This is interesting, because it offers the possibility of a consistency 
check on the inflationary hypothesis which is {\em independent} of the 
choice of inflationary potential.

The three inflationary parameters for large-scale structure are
\begin{enumerate}
\item The amplitude of density perturbations, $\delta_{{\rm H}}$, defined 
below.
\item The spectral index of the density perturbations $n$.
\item The relative contribution of gravitational waves to large-angle CMB 
anisotropies, $R \equiv C_{\ell}^{{\rm GW}}/C_{\ell}^{{\rm DP}}$, where 
$\ell$ could for instance be taken to be 10, the {\it COBE} pivot scale.
\end{enumerate}
The equation describing the redundancy is
\begin{equation}
R \simeq - 2 \pi n_{{\rm GW}} \,,
\end{equation}
where $n_{{\rm GW}}$ is the spectral index of the gravitational waves. 
Although a very distinctive signature of inflation, it seems very unlikely 
that one will ever be able to carry out this test since $n_{{\rm GW}}$ is 
almost certainly impossible to measure. By contrast, one should be extremely 
optimistic about measuring the non-degenerate parameters $n$ and $R$.

One sometimes sees the relation $R \simeq 2 \pi (1-n)$ quoted as an 
inflationary signature. This is in fact the prediction of a specific 
model, the power-law inflation model (which always gives $n<1$), and is not 
general enough to describe generic slow-roll inflation, for which $n$ and 
$R$ enter independently at the same order in the slow-roll expansion. It is 
quite a useful relation nevertheless, because for $n<1$ it gives about the 
largest amount of gravitational waves one finds in any inflation model, and 
so a good range to consider is $R$ going from 0 up to this value. However, 
for purposes such as fitting CMB anisotropies, there seems little purpose in 
adopting such a relation, because simply by including $n$ and $R$ 
independently one is looking at the generic slow-roll inflation situation. 
Since one is anyway fitting for a large number of cosmological parameters, 
the addition of one more can hardly degrade the fit at all and has the 
enormous benefit of generality.

If one has a particularly complex inflationary potential, one might need 
more parameters to describe its influence. This can be achieved by a 
technique known as the slow-roll expansion. Up to a point, having to 
introduce extra parameters is actually a good thing, because if you feel you 
need them to describe the data, that means that potentially there is more 
information, in the form of extra parameters, to be learnt about inflation 
from the observations. Since there are a lot of parameters anyway, a couple 
more shouldn't particularly degrade the fit. Equally though, it would be 
possible to have too much of a good thing and I can't imagine that anyone 
would be pleased if it was thought that inflation might add say twenty new 
parameters. Fortunately that will not be the case\footnote{Though it is 
possible to have models, such as double inflation, that have features so 
drastic that they can't be described within this perturbative framework at 
all.}.

\section{Cosmological parameters}

When one considers short-scale information, such as large-scale structure 
data, the theoretical prediction depends on the whole range of cosmological 
parameters, including

1.~The Hubble parameter $h$.

2.~The total matter density $\Omega_0$.

3.~The cosmological constant $\Lambda$.

4.~The baryon density $\Omega_{{\rm B}}$.

5.~The hot dark matter (HDM) density $\Omega_{\nu}$. (It is always 
assumed there'll be some CDM).

6.~The effective number (at the photon temperature) of massless species 
$g_*$.

7.~The redshift of reionization $z_{{\rm R}}$.\\
Further complexity can be introduced in many ways, for example by allowing 
the dark matter to decay, by allowing the HDM to be composed of more than 
one particle type or by allowing the hot component to violate the usual 
mass--density relation.

All of these parameters have effects on various kinds of observations, and 
coupled with the three input parameters from inflation give a fairly large 
number of parameters to be dealt with. The sort of data a new generation CMB 
anisotropy satellite might produce could quite conceivably just 
simultaneously fit all, or a sizeable subset, of them, but present data are 
a long way off and it is usual to work within some simplifying assumptions. 
Our analyses have been unusual in that they attempt to retain all the 
inflationary parameters, as well as a reasonable number of the cosmological 
ones.

Although the list of cosmological parameters looks intimidating, combining 
all the options seems very unattractive. Also, various scalings can be 
utilized, since many changes have an effect very similar to changing the 
Hubble parameter (the main effect of which is to shift the epoch of 
matter--radiation equality), which allows many options to be considered at 
once. The epoch of reionization is only important if one considers 
intermediate-scale CMB anisotropies. 

\section{The {\it COBE} normalization}

More than anything else, this section is a request to those carrying out 
fittings to the {\it COBE} data to quote the best-fit amplitude of the 
{\em matter} power spectrum, as well as radiation anisotropy quantities such 
as $Q_{{\rm rms}}$. In general it is far from trivial to get from one to the 
other. It doesn't really matter in what form this is done, but I'd like to 
put in a word in favour of a quantity we call $\delta_{{\rm H}}$. 

When the spectrum of density perturbations ${\cal P}$ is defined so that the 
variance is
\begin{equation}
\sigma^2(R) = \int_0^{\infty} W^2(kR) \, {\cal P}(k) \, \frac{{\rm d}k}{k}
	\,,
\end{equation}
where $W(kR)$ is whatever window function does the smoothing, then the 
spectrum can be broken up into pieces as
\begin{equation}
{\cal P}(k) = \left( \frac{k}{aH} \right)^4 \, \delta_{{\rm H}}^2(k)
	\, T^2(k,t) \, \frac{g^2(\Omega)}{g^2(\Omega_0)} \,.
\end{equation}
The final term is the usual growth suppression factor, which can be applied 
at any epoch provided the $\Omega$ at that time is used. This term vanishes 
if $\Omega = 1$. The transfer function is normalized to unity on large 
scales. In general it may be time dependent, for example if there is an HDM 
component, but in a CDM universe it is time independent. The very first term 
carries the remaining time dependence, that of a critical density CDM 
universe. 

Finally then, $\delta_{{\rm H}}$ is the initial spectrum of perturbations. 
It is time-independent, and amounts (when multiplied by the growth 
suppression term) to a proper definition of what is 
meant by the perturbation at horizon crossing. In the special case $n = 1$ 
it is also $k$-independent. One can specify a fit to the matter power 
spectrum by giving its value at $k=a_0 H_0$, the present Hubble radius. For 
the four-year {\it COBE} data, one finds for CDM models with a cosmological 
constant$^{3)}$ that
\begin{equation}
\delta_{{\rm H}}(a_0 H_0) = 1.94 \times 10^{-5} \, \Omega_0^{-0.785-0.05 
	\ln \Omega_0} \, \exp \left[ f(n) \right]
\end{equation}
where 
\begin{eqnarray}
f(n) & = & - 0.95 (n-1) - 0.169 (n-1)^2 \quad 
	\mbox{No gravitational waves}\\
   & = & \; 1.00 (n-1) + 1.97 (n-1)^2 \quad \; \; \mbox{Power-law inflation}
\end{eqnarray}
A similar fitting function can also be found for the open universe case. The 
joy of this formula is that it says everything about the {\it COBE} 
normalization in a single fitting formula. To a good approximation, this fit 
is {\em independent} of $h$, $\Omega_{{\rm B}}$ and the nature of the dark 
matter.

\begin{figure}[t]
\centering
\leavevmode\epsfysize=10.5cm \epsfbox{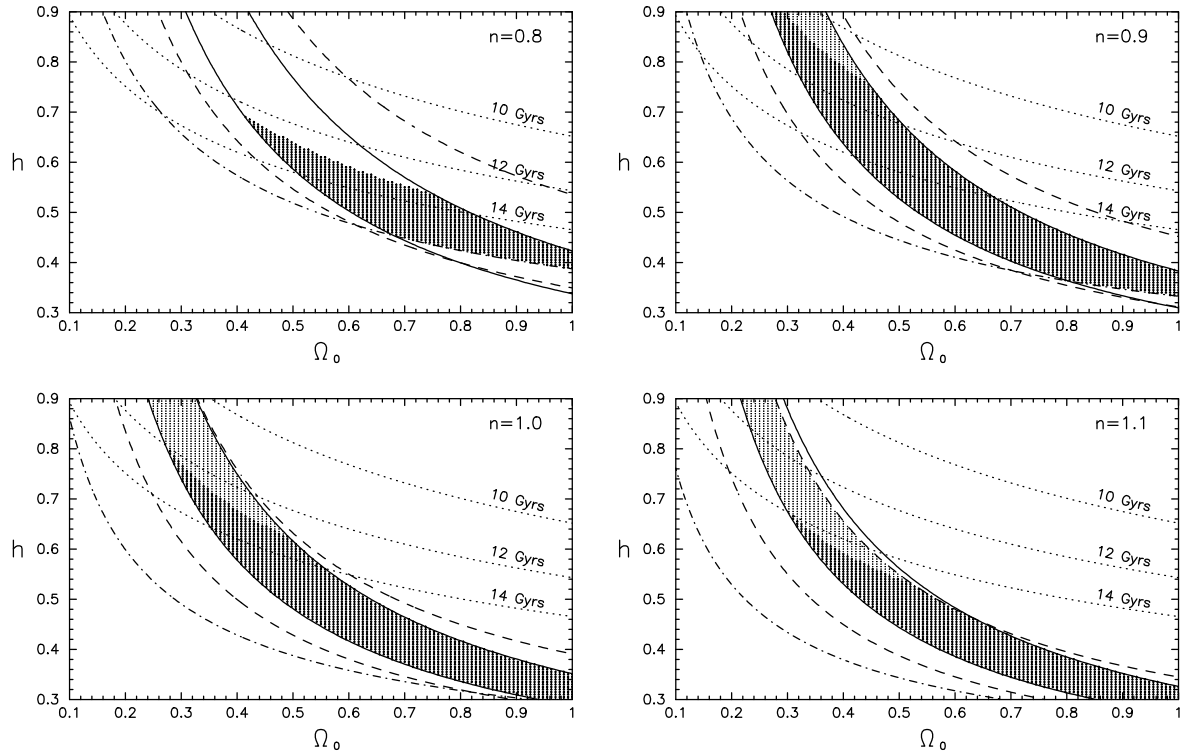}\\
\parbox{16cm}{\caption[figure1]{\small 
\baselineskip 12pt 
Varying $n$, but with no gravitational waves, for 
spatially flat CDM models. All constraints are plotted at 95 per cent 
confidence, and models are normalized to the 4 year {\it COBE} data. The 
constraints are galaxy clustering (solid), cluster abundance (dashed) and 
DLAS abundance (dot-dashed). Bulk flows and quasar abundance are less 
constraining. Contours of constant age are shown as dotted lines. The 
allowed region is shown with two different shadings, both highlighting the 
parameter space not excluded at more than 95 per cent confidence on any 
single piece of data. The lighter shading shows models where the optical 
galaxies have to be antibiased at $8 h^{-1}$ Mpc. Finally, the unshaded 
region in the $n=0.8$ plot which is allowed by all plotted data is excluded 
by Doppler/acoustic peak height.}}
\end{figure}

\section{CDM models with a cosmological constant}

Our strategy has been to assemble observational data which can be 
interpreted in terms of linear and quasi-linear theory. This allows a rapid 
assessment of which regions in parameter space are allowed or disallowed. 
The observations considered are large-angle CMB anisotropies ({\it COBE}), 
intermediate angle CMB anisotropies, bulk motions, galaxy clustering and the 
abundances of galaxy clusters, quasars and damped Lyman alpha systems 
(DLAS). Extra considerations that can be brought into play are the age of 
the universe, suppression of intermediate angle anisotropies by reionization 
and the question of whether (optically identified) galaxies are permitted to 
be anti-biased.

Figure 1 shows the sort of constraints that can be applied to CDM models 
in spatially flat universes. These plots show tilt $n$, but not 
gravitational waves. There is a sizeable allowed region for each $n$ shown, 
so observations do not really constrain the inflationary parameters. I'll 
just make some brief points about these plots.

The age problem is usually cited as motivation for going to cosmological 
constant models. However, within the region fitting large-scale structure 
observations, the low density models are actually younger than the high 
density ones. Of course, as $\Omega_0$ approaches one the required $h$ is 
alarmingly low, and one might feel inclined to introduce a hot component, 
which permits larger $h$ with critical density.

In the low density region, optical galaxies typically have to be 
anti-biased (and of course {\it IRAS} galaxies more so). This is thought 
unlikely, though it's not clear exactly what observations this is 
supposed to be in conflict with. For $n=0.8$, there is a region which is 
allowed by all data {\em except} intermediate-scale CMB anisotropies; in 
that region of parameter space there isn't really a Doppler/acoustic peak at 
all. This is a sign of things to come; intermediate CMB anisotropies have 
the potential to exclude swathes of parameter space in the future.

\section{Conclusions}

Cosmologists are beginning to take seriously the possibility that one can 
determine the whole range of cosmological parameters. Within that context, 
one appreciates that it is possible to also include information from 
inflation, and attempt to fit for inflationary parameters at the same time 
as the cosmological parameters. The most popular inflationary paradigm, the 
slow-roll approximation, only introduces two extra parameters ($n$ and $R$) 
that one didn't have to consider anyway, and there is good reason to be 
optimistic that one can constrain these. 

However, once one takes the extra inflationary input into account, it is 
clear that present observational data fall some way short of providing any 
telling constraints. We have found that it is possible to get an adequate 
fit to present data within almost any context. There are viable regions of 
parameter space for
\begin{itemize}
\item {\bf CDM models$^{2)}$:} Requires some or all of low $h$, high 
$\Omega_{{\rm B}}$ or tilt to $n <1$. Gravitational waves don't help much, 
but they are not very strongly constrained. It is however very hard to fit 
the data for $h \geq 0.50$. Adding extra massless species or decaying dark 
matter will also work though we haven't investigated them in detail 
ourselves.
\item {\bf CHDM models$^{2)}$:} The same general picture as CDM models, but 
allows a higher value of $h$, at least up to 0.6, provided the amount of hot 
dark matter is chosen wisely.
\item {\bf Low density CDM$^{1,3)}$:} Can be made to work either in the open 
case or in the flat case with a cosmological constant. Observationally, no 
strong preference between the open and flat cases.
\end{itemize}
This situation should not remain for long. We stand at a tantalizing time, 
where observations are just good enough to exclude the more extreme 
inflationary models. We can look forward in the near future to a time when 
inflationary and cosmological parameters are extremely well determined, at 
which point we can expect most inflation models to be ruled out. Or maybe 
even all!

\vspace*{12pt}
\noindent
{\bf Acknowledgments:} I thank my collaborators on work described herein, 
namely David Lyth, Dave Roberts, Bob Schaefer, Qaisar Shafi, Pedro Viana and 
Martin White. I also thank Pedro (again!) for producing the figure. I am 
supported by the Royal Society and acknowledge use of the Starlink computer 
system at the University of Sussex. 

%%%%%%%%%%%%%%%%%%%%%%%%%%%%%%%%%%%%%%%%%%%%%%%%%%%%%%%%%%%%%%%%%%%%%%
\frenchspacing
%%%%%%%%%%%%%%%%%%%%%%%%%%%%%%%%%%%%%%%%%%%%%%%%%%%%%%%%%%%%%%%%%%%%%%
\section*{References}
I regret there has been no space here to provide an adequate reference list. 
Full references can be found in the papers cited here; apologies to the vast 
number of people who are missing out!

\vspace*{12pt}

{\baselineskip 16pt
\noindent
1.~A. R. Liddle, D. H. Lyth, D. Roberts and P. T. P. Viana, Mon. Not.
	Roy. Astr. Soc. {\bf 278}, 644 (1996).\\
2.~A. R. Liddle, D. H. Lyth, R. K. Schaefer, Q. Shafi and P. T. P. Viana,
	Mon. Not. Roy. Astr. Soc., to appear (1996), astro-ph/9511057.\\
3.~A. R. Liddle, D. H. Lyth, P. T. P. Viana and M. White, Mon. Not. Roy.
	Astr. Soc., to appear (1996), astro-ph/9512102.
}
%%%%%%%%%%%%%%%%%%%%%%%%%%%%%%%%%%%%%%%%%%%%%%%%%%%%%%%%%%%%%%%%%%%%%%
\end{document}